\documentclass[a4paper,article,10pt]{memoir}
\usepackage[english]{babel}
\usepackage{kaeiog}
\usepackage{dcolumn}
\usepackage{bm}
\newfont{\mm}{msbm10}
\def\square{\hbox{\vrule\vbox{\hrule\phantom{l}\hrule}\vrule}}
\begin{document}
%
%
%
\title{
Breakthrough in Interval Data Fitting\\
I.~The Role of Hausdorff Distance}
%
%
\author{%
    Marek W.~Gutowski\\
\bigskip\small
Institute of Physics, Polish Academy of Sciences, 02--668 Warszawa, Poland,
email:~{gutow@ifpan.edu.pl}
}
%
%
\maketitle
\thispagestyle{empty} 
%
%
%
    \begin{abstract}
~---~This is the first of two papers describing the process of fitting experimental data
under interval uncertainty.\ 
Probably the most often encountered use of \textbf{global optimization methods\/}
is finding the so called {\em best fitted\/} values of various parameters, as well as
their uncertainties, based on experimental data.\  
Here I~present the methodology, designed from the very beginning
as an interval-oriented tool, meant to replace to the large extent the famous
\textbf{Least Squares (LSQ)\/} and other slightly less popular methods.\
Contrary to its classical counterparts, the presented method does not require any
poorly justified prior assumptions, like {\em smallness\/} of experimental uncertainties
or their normal (Gaussian) distribution.\  Using interval approach, we are able to fit
rigorously and reliably not only the simple functional dependencies, with no extra effort
when both variables are uncertain, but also the cases when the constitutive equation
exists in \textbf{implicit} rather than \textbf{explicit functional form}.\  The magic word
and a~key to success of \textbf{interval approach\/} appears the \textbf{Hausdorff
distance}.
    \end{abstract}

\CCLsection{Introduction}
Handling experimental uncertainties lies at the heart of physics and other sciences.\
The situation is rather clear during simple measurements, for example when determining
the mass, $m$, of an object under investigations.\ In this case the uncertainty of a single
measurement is equal to one half of the smallest division ($\Delta$) of the measuring
device, here the balance.\  Consequently, we can be sure that the unknown mass
certainly belongs to the interval $\left[m-\Delta/2,\, m+\Delta/2\right]$, where $m$ 
is a~direct readout from the instrument's scale.\  We can try to increase our confidence
into the result by repeating the measurement several times or by using several
measuring devices, maybe delivering the results with different precision $\Delta$.\ 
By doing so, we necessarily switch to the probabilistic way of presenting measurement's
results, namely as a~pair of two numbers: the most likely value, $m$,  and its standard
deviation, $\sigma(m)$, expressing the uncertainty of a~measurement.\

Nowadays more and more measurements belong to a~class called {\em indirect
measurements\/}, where we deduce the numerical values of various interesting
parameters without comparing them directly to the appropriate standard.\
When determining the resistance of a~sample we usually take a~series of
measurements, obtaining many pairs (voltage, current).\  Then, applying the Ohm's
law relating those two physical quantities, we find the single proportionality constant,
called resistance, $R=\textrm{voltage}/\textrm{current}$ (or its inverse: conductance).\
Using many measurements, not just one, we switch naturally to the probabilistic
form of result's presentation: as the most likely value and its standard deviation
(here: $R$ and $\sigma(R)$).

This situation is highly unsatisfactory.\
First, we can never be $100$\% sure whether the partial uncertainties are indeed
{\em small\/}, thus justifying the application of various `error propagation laws'.\
Secondly, neither the probability density function of measurements nor of the result
need not to be normal (Gaussian), thus invalidating the frequent claim that the
probability of the true value belonging to the interval $\left[m-\sigma,\, m+\sigma\right]$
is approximately equal to $67$\%.\  In reality, the overwhelming majority of contemporary
measuring devices deliver discrete, digital results.\  Such results obviously cannot
be normally distributed.\

In this paper I present other ways of experimental data processing, using 
interval methods.\  Here we operate on guaranteed rather than probabilistic
quantities, thus the results obtained on this way should also be guaranteed.\
In practice, the results produced by straightforward interval implementation
of the classical methods often appear disappointing.\  In this paper I point
to the possible reasons and show a~remedy.

\CCLsection{Basics of interval calculus}
From now on we will use {\em intervals\/} every time our numbers are uncertain.\
Specifically, by interval ${\mathbf x}$ we will understand a (sub)set of real
numbers:
\begin{equation}
{\mathbf x} = \left[\underline{x}, \overline{x}\right] = \left\{\hbox{\mm R}^{1}\ni\,x:\ \underline{x}
\leqslant x \leqslant \overline{x} \right\}.
\end{equation}
The set of all intervals is usually written as \hbox{\mm IR}.\ The familiar real
numbers can be identified with intervals of type ${\mathbf x}=[x,x]$, i.e. having their
{\em lower\/} and {\em upper bound\/} equal to each other, and called {\em degenerate
intervals\/}, {\em thin intervals}, or {\em singletons\/}.\  This way the set of intervals
may be regarded to be an extension of real numbers:\
$\hbox{\mm R}^{1} \subset \hbox{\mm IR}$.

\medskip
 It is possible to define the arithmetic operations on intervals:

\begin{description}
\item [addition:]\ ${\mathbf x} + {\mathbf y} = \left[\underline{x} + \underline{y},\
\overline{x} + \overline{y}\right]$;
\item [subtraction:]\ ${\mathbf x} - {\mathbf y} = \left[\underline{x} - \overline{y},\ 
\overline{x} - \underline{y}\right]$;
\item [multiplication:]\
${\mathbf x} \cdot {\mathbf y} = \left[\min Z,\ \max Z\right]$,\
where $Z = \left\{\underline{x}\underline{y},\ \underline{x}\overline{y},\
\overline{x}\underline{y},\ \overline{x}\overline{y}\right\}$;
\item [division:]\
${\mathbf x}/{\mathbf y} = {\mathbf x} \cdot \frac{1}{\mathbf y},$\
where\ $\frac{1}{\mathbf y}=\left[1/\overline{y},\ 1/\underline{y}\right],\
 \quad 0\notin{\mathbf y}$.
\end{description}

\medskip
One can understand the right hand sides of the above definitions as intervals
{\em certainly\/} containing the results of the corresponding operations when the
first operand is arbitrarily chosen from interval ${\mathbf x}$ and the second -- from
interval ${\mathbf y}$.\ One can also easily check that we recover precisely the
ordinary results whenever\ $\underline{x}=\overline{x}$\ and\ $\underline{y}=\overline{y}$.\
It has to be stressed, however, that in computer calculations it is necessary to properly
round every result.\  The so called {\em outward\/} rounding is appropriate here,
that is the lower bound ($\underline{x}$) should be always rounded towards $-\infty$
and the upper bound ($\overline{x}$) should be always rounded towards $+\infty$.\
Only then we can be sure that our results are certain (guaranteed), i.e. containing
the true result with probability~$1$.\  A warning is in place here: never
treat the interval as a~random variable uniformly distributed in its interior.\  If this
was so, then the sum of two intervals would have to represent a~triangularly distributed
random variable, what is -- obviously -- not the case.

\medskip
Since intervals are sets, then it is natural to operate on them as on sets, not only
as on numbers.\ Consequently we can compute the intersection of two intervals
(it may happen to be an empty set, conveniently written as $[+\infty, -\infty]$ --
an illegal interval), or their set-theoretical sum (union).\ The union of two non-empty
intervals is never empty but not necessarily is an interval again.\  This may be
sometimes troublesome, so it is desirable to introduce the notion of
{\em convex hull\/}, or shortly {\em hull\/} of two intervals as a~smallest interval
containing them both:
\begin{equation}
\hbox{\textrm hull}\left({\mathbf x}, {\mathbf y}\right)\,\equiv\,\square \left({\mathbf x},
{\mathbf y}\right) = \left[ \min (\underline{x}, \underline{y}),\
\max (\overline{x}, \overline{y}) \right]
\end{equation}

It is remarkable that the finite number of arithmetic operations (performed on intervals'
endpoints) is all we need to process uncountably infinite subsets of real numbers.\
The nice feature of interval computations is its straightforward extensibility to operate
on vectors, matrices, or real-valued functions as well.

\medskip
The essence of interval calculations is to deliver guaranteed results, but sometimes we
obtain intervals which are wider\footnote{The width of interval is a~positive real number: 
w$({\mathbf x})=\overline{x}-\underline{x}$.} than necessary.\
The {\em overestimation\/} is usually hard to asses (and may be quite large!) but will
happen almost surely when in a given expression any variable appears more than
once; for example:
$[0, a]-[0,a] = [-a, a]$\ and \underline{not}\ $0=[0,0]$\ --- as one might expect.\  This
phenomenon is called a {\em dependency problem\/}.  In particular, we have:
\begin{equation}
{\bold x}\left({\bold y} + {\bold z}\right) \subseteq\ {\bold x}{\bold y} + {\bold x}{\bold z}
\end{equation}
So, the order of calculations matters.\ Fortunately, the overestimation usually vanishes
when the widths of interval operands go to zero.\  Exceptions do exist but it is not easy to present a~practical example.

\medskip
The vectors with interval components are usually called {\em boxes} and it is easy to
see why.\  For more information on interval computations the reader is referred to
other sources \cite{website}.

\CCLsubsection{Common features of interval algorithms}

Interval-oriented algorithms usually operate on lists of boxes and treat them accordingly
to their current status.\  {\em Good\/} boxes are retained for further reference, while {\em
bad\/} boxes are discarded as soon as possible.\  The third category, called
{\em pending\/}, {\em unresolved\/} or {\em undetermined\/}, is the most interesting
and as such is the main object of processing.\  The general idea is to start from a~single
big box, suspected to contain the solution (or solutions) and therefore initially labelled as
pending.\  Pending boxes, one by one, are divided into smaller parts (usually two) and
disappear from the list.\ Offspring boxes are subjected to one or more tests aiming at determining their current status.\  Some of them are recycled back onto pending list,
while {\em bad\/} are immediately discarded, and {\em good\/} are collected separately
as a~part of the final result.\  The procedure terminates when either the list of pending
boxes is empty or contains only `small' boxes.\  On exit, the result is a union of good
boxes, possibly appended with pending ones.\  Empty output is a~\textbf{proof} that
the solution(s) of our problem, if any, are located outside the initial box.

Of course, the exact meaning of {\em bad\/} and {\em good\/} depends on context.

\CCLsection{The problem of experimental data fitting and its solution}
Suppose we have a theory ${\mathcal T}$, characterized by $k$ unknown
parameters ${\mathbf p}_{1}, \ldots, {\mathbf p}_{k}$, and $N$ ($N\,\ge\,k$)
results of measurements\ ${\mathbf m}_{1}, \ldots, {\mathbf m}_{N}$, each taken
at different values of some well controlled variables, for example the varying 
temperature or magnetic field.\  Let ${\mathbf x}_{j}$ denotes the value(s) of controlled
variable(s) (environment) during the measurement ${\mathbf m}_{j}$.\ In what follows
we will use the simplified notation:

\begin{equation}
\begin{array}{lcl}
{\mathcal T}({\mathbf p}_{1}, \ldots, {\mathbf p}_{k}; {\mathbf x}_{j}) &=& {\mathbf t}_{j},
\quad\ j = 1, \ldots, N\\
{\mathbf m}_{j}({\mathbf x}_{j}) &\equiv& {\mathbf m}_{j},
\label{shorthand}
\end{array}
\end{equation}

where the upper line describes the theoretical outcomes of the experiment and the lower one
-- actual experimental results.\  Usually each theoretical outcome ${\mathbf t}_{j}$
is crisp\footnote{We call the object {\em crisp\/} (point, point-wise, point-like)
in contrast to the one having interval character, be it a real number, vector or matrix.}
when all the arguments of\ ${\mathcal T}$\ (including ${\mathbf x}_{j}$) are crisp.\
Contrary, the measurements ${\mathbf m}_{j}$ are always uncertain and will be
considered from now on to be intervals (or more generally: interval vectors).

The task for an experimentalist is to adjust the values of all unknown parameters
${\mathbf p}_{1}, \ldots, {\mathbf p}_{k}$,\  in such a way that every measurement
${\mathbf m}_{j}$ differs as little as possible from the corresponding theoretical
prediction ${\mathbf t}_{j}$.\  In fact, we would be happy to finish the
procedure observing
\begin{equation}
{\mathbf t}_{j}\, \cap\,{\mathbf m}_{j}\,\ne\,\varnothing,\quad\,\forall j\,\in\,\{1, 2, \ldots, N\}
\label{intersect}
\end{equation}
(in simple words: the theoretical curve passes through all the experimental points)\
and with unknown parameters,\ ${\mathbf p}$'s,\  as narrow as possible.

Note that the requirement ${\mathbf t}_{j}\,\subseteq\,{\mathbf m}_{j}$ for every $j$,
although tempting,  is going too far: our results (if they exist) would be severely
underestimated.\  Yes, we want to be accurate, but we can't afford to tolerate
underestimates.

It is out of scope of this article to elaborate the details of many existing particular
procedures \cite{Jaulin-IntervalComp94}
--\cite{SCAN2008}
 aimed to maximally
contract the initial box of unknown parameters ${\mathbf p}_{1}\times\cdots
\times{\mathbf p}_{k}$ in such a~way that all the relations
(\ref{intersect}) are satisfied.\   Some of those methods (see \cite{ja}) deliver a single box
being an interval hull of solutions, while others, more time-consuming, output
more boxes, covering with certainty the solution set.\ 
Technically: we call a box {\em bad\/} when at least one of inequalities (\ref{intersect})
is violated at all its internal points and {\em pending\/} otherwise.\  In rare cases it happens
to encounter\  ${\mathbf t}_{j}\subseteq\,{\mathbf m}_{j}$\ for all\  $j$'s -- such box
is {\em good\/} and requires no further processing.

\CCLsubsection{Advantages and disadvantages of the rigorous solution}
\label{difficulties}

The practical implementations of the above described approach are still
rare \cite{JPhysA30-7733, swim08-aufray} and one may wonder why.\
The unquestionable mathematical rigor of a~procedure is certainly among
its major advantages.\  Unfortunately, for very practical reasons, this is also its
weak point.\ And here is why:
\begin{itemize}
\item
the routine requires guaranteed intervals as measurements,\, i.e. the ones
containing true outcomes of an experiment with probability equal exactly to $1$.\
It is impossible to satisfy this requirement in laboratory practice since measurements
usually have the form $m=m_{0}\pm\sigma(m)$, where $m_{0}$ is a mean value
and $\sigma(m)$ -- its standard deviation.\  Nobody knows (or cares) what is the
distribution of $m$, even its support is usually unknown (well, some physical quantities,
like mass or absolute temperature, have to be non-negative).\  Consequently, taking intervals
$\left[m_{0}-\sigma,\, m_{0}+\sigma \right]$ as input data, we will most likely finish with
empty set of results.\  This is because at least one-third of $N$ relations
(\ref{intersect}) has to fail, for arbitrary (even crisp!) set of unknown parameters
$\left\{p_{1}, \ldots, p_{k}\right\}$;

\item
the quick and dirty fix, coming to the mind in the above situation, is to use wider
intervals, like\ $\left[m_{0}-3\sigma,\, m_{0}+3\sigma\right]$,\ as `almost guaranteed' input data.\
Depending on the individual luck, this trick may or may not work.\  If it doesn't then we
are left without even a~foggy idea what are the values of our unknown parameters.\
If it works then uncertainties of so obtained parameters often seem exceedingly large,
very pessimistic, at least when compared with those obtained on other ways.\  Besides:
are we entitled to `scale back' the obtained uncertainties dividing them all by~$3$?
The only honest answer is {\em no\/};

\item
sometimes, when the fix works, the uncertainties of unknown parameters appear
unrealistically, not to say suspiciously, small.\  This will surely happen when among our
data there is at least one element which should be labeled as a `near outlier'.\  It may
be due to the undetected data transmission/recording error, power line fluctuations, or whatever.\

\item
another approach is to require fitted curve to pass through at least\ $N_{1}<N$\ experimental
points, without any prior indication which points are to be preferred.\  The sensible choice
for $N_{1}$ is $N_{1}\approx\,2N/3$.\  We will not discuss further this idea.
\end{itemize}

\medskip
On the other hand, when all our data are credible but the solution set is empty
-- our theory $\mathcal{T}$ must be flawed.\  This could be a very strong statement,
but is not.\  There is one more possibility: the experimental uncertainties are
seriously underestimated, deliberately or otherwise.\  So, experimenters, be warned:
cheating will be severely punished by interval methods and will bring you nowhere!

\medskip
Concluding:  while the rigorous approach has many advantages, we definitely need
something else.\

\CCLsubsection{Back to the basics}
The situation is especially upsetting when experimental data look fine,
and we are sure the theory is correct, yet the interval routine returns
no answer at all.\  What can we do?

In classical data analysis, the LSQ method (\textbf{L}east \textbf{Sq}uares)
is most commonly used to find unknown parameters.\  In short, its essence is to minimize
the quantity called {\em chi squared\/}:
\begin{equation}
\chi^{2}({\mathbf p}) = \sum_{j=1}^{N} \frac{\left(m_{j}-t_{j}({\mathbf p})\right)^{2}}{\sigma_{j}^{2}},
\label{chi2}
\end{equation}
where ${\mathbf p}$ is a (crisp) vector of $k$ unknown parameters -- arguments of the
function $\chi^{2}$.\  We are looking for such a~vector 
${\mathbf p}^{\star}$ as to have $\chi^{2}({\mathbf p}^{\star})=\min$.\  It is well
known that LSQ method never fails and always produces some results,
even for completely wrong theory.\

\medskip
The LSQ method is due to Carl Friedrich Gau\ss\  (1777--1855) and was originally
invented by him around 1794.\   Later on, in 1809, the same author gave it solid
statistical interpretation.\  We will not proceed with statistical properties of $\chi^{2}$
and LSQ method in general, instead we will rather concentrate on the original idea.
And it was like that:

\medskip
Due to unavoidable experimental uncertainties (then called {\em errors\/}), it is unlikely
to draw a line representing theoretical outcomes as a function of environment ($x$) and
have the experimental points $m_{j}(x)$ lying precisely on this curve at the same time.\
Note that the notion of standard deviation of the measurements was practically unused
those days, so the measurements were just numbers.\  This means there were no denominators
in formula (\ref{chi2}).\  Varying the unknown parameters $p_{j}$'s, we deform the theoretical
curve so as to have it running as closely to the experimental points as possible, thus
solving the problem.
\medskip

\noindent\textbf{Remark:}
\begin{quote}
Gau\ss\ could not speak about the distance (between a~theoretical curve and experimental
points), at least not in a~strict
mathematical sense, since the notion of {\em distance\/}  was introduced into
mathematics many years later, by another German mathematician, Felix Hausdorff
(1868--1942), and, independently, in 1905, by Romanian mathematician Dimitrie Pompeiu
(1873--1954) \cite{pompifip}.
\end{quote}

\medskip
Having known the notion of distance, Gau\ss\  would almost certainly use it in his
early idea of LSQ method.\  Indeed, the formula (\ref{chi2}) is nothing else as squared
Euclidean distance in $N$-dimensional space, where $\sigma_{j}$ serves as
a~unit length in direction $j$.

\CCLsection{Interval version of LSQ method}
The problem of global optimization is well studied.\  Interval researchers have also
contributed their share to this field.\ They have already devised many interval-thinking-inspired, rigorous procedures aiming to solve such tasks.\  Minimizing
$\chi^{2}$ is obviously one of them.\  Simple optimization cases, in LSQ sense, were investigated since at least 1990 \cite{LeastSquares1900}.\ Several closely related
problems have been successfully solved using rigorous interval algorithms, yet the
specific difficulties, recalled in  Sec.~\ref{difficulties}, still have not been addressed
satisfactorily.\ 
 
The enormous potential of interval methods for uncertain data processing has been
recognized some time ago (1993) by Walster and Kreinovich \cite{ACM-SIGNUM-Newsletter-vol31-iss2}.\
Kosheleva and Kreinovich (1999) pointed that the cost of interval approach only
slightly exceeds the traditional (probabilistic) one \cite{RC5-81}.\
Mu\~noz and Kearfott (2001) have shown that non-smooth cases should be essentially
no more difficult than regular ones \cite{2001robustness}.\
Only one year later (2002), Yang and Kearfott heralded `new paradigm' and `new way
of thinking about data fitting' \cite{2002SIAM}.  Unfortunately, they did not show how to practically implement their ideas other than for linear equation set.\  
Finally, (2005) Zhilin \cite{RC11-433} found the solution for the case when measurements
are (multi)linearly related to the environment.\

\CCLsubsection{Correct interval form of $\chi^{2}$}

How to `translate' the fundamental formula (\ref{chi2}) into its interval counterpart?\
The first idea is to simply replace real-valued measurements $m_{j}$ by their interval
representations ${\mathbf m}_{j}$,\ apply the same trick to theoretical predictions
$t_{j}\,\rightarrow\,{\mathbf t}_{j}$,\ and retain real-valued $\sigma_{j}$'s.\ Consequently,
the functional $\boldsymbol{\chi}^{2}$ becomes interval-valued as well.\  So far, so good,
but wait.\

\medskip
In the light of what was said before, we have to rewrite  (\ref{chi2})
in terms of distance between experimental and theoretical (predicted, simulated)
points.\  Yet the expression ${\mathbf m}_{j} - {\mathbf t}_{j}$, nor even
$\left|{\mathbf m}_{j} - {\mathbf t}_{j} \right|$, is {\em not} the correct
mathematical distance!\  There are two reasons for that:
\begin{itemize}
\item it is not a real number, and
\item the implication $\left({\mathbf m} = {\mathbf t}\right)\,\Rightarrow\,
\left({\mathbf m} - {\mathbf t} = 0\,\right)$ is false.
\end{itemize}

This naturally raises the question: what {\em is\/} the distance between two intervals, here
$\mathbf{m}$ and $\mathbf{t}$?  Following Hausdorff, we will use his metrics, adapted
for {\mm IR} space by Moore \cite{Moore} as:
\begin{equation}
d\,({\mathbf a},\,{\mathbf b}) = \max\,\left(\left|\,\underline{a} - \underline{b}\,\right|,\,
\left|\,\overline{a} - \overline{b}\,\right|\,\right).
\label{Hdist}
\end{equation}
Before continuing, let us only note that another function
 $d^{\prime}\,({\mathbf a},\,{\mathbf b}) = C\cdot\,d\,({\mathbf a},\,{\mathbf b})$,
where $C>0$ is a fixed number, is a correct distance, too.\
Additionally, the mathematical distance is not expressed neither in miles nor millimeters, or
in any other units -- it is simply dimensionless.\  This is important, since using the strict
mathematical distance between intervals we will be able to fit our experimental data
obtained from two (or more) completely different experiments, provided the sets
of unknown parameters, relevant to each experiment separately, have non-empty
intersection.\ Simply speaking -- it is possible to fit several curves simultaneously,
in a~single run.

\medskip\noindent
\textbf{Remarks:}
\begin{quote}
The Euclidean distance between theoretical and experimental points is not the only
thinkable distance.\  We will discuss other $N$-dimensional metrics in the other article.\
We will stick, however, to the Moore-Hausdorff distance as one-dimensional metrics
in {\mm IR}, no matter that there are other choices, see for example \cite{Palumbo689}
and \cite{PattRecogn}.\  Note also that Moore-Hausdorff distance specialized to the real
line simplifies to the familiar form: $d(a,b)\,=\,\left|\,a - b\,\right|$,\ as one might expect.
\end{quote}

\CCLsubsection{Analyzing $\chi^{2}$ term by term}
Consider now again a~single term in (\ref{chi2}).\  What is the distance between the
predicted value ${\mathbf t}$ and the true value $m^{\star}$\ ($m^{\star}$ is a~real
number, not interval)?\
All we know about $m^{\star}$ is that it satisfies the double inequality:\ $\underline{m}
\leqslant m^{\star} \leqslant \overline{m}$, with exact value of $m^{\star}$ remaining
unknown.\  According to the definition (\ref{Hdist}) we have ($m^{\star} = \left[m^{\star}, 
 m^{\star}\right] = {\mathbf m}^{\star}$):
\begin{equation}
d\,({\mathbf t},\,m^{\star}) = \max \left( \left| \underline{t} - m^{\star} \right|,\,
 \left| \overline{t} - m^{\star} \right|\right)\,\in\,\protect{\hbox{\mm R}}
\label{H-point}
\end{equation}
We don't know which internal point of ${\mathbf m}$ is equal to $m^{\star}$.  Suppose for
a moment that $m^{\prime}$ coincides\footnote{For a moment we will be dealing with some
$m^{\prime}$, not even necessarily contained in $[\underline{m}, \overline{m}]$,
rather than with $m^{\star}$.\  This is just for purity: $m^{\star}$ is a fixed number and
we want $m^{\prime}$ to be variable.}
with one of the endpoints of ${\mathbf m}$, say $m^{\prime}=\underline{m}$.\
Call the current distance $d({\mathbf t}, m^{\prime}=\underline{m})=\xi$.\   What happens
when $m^{\prime}$ gradually moves to $\overline{m}$ -- the other endpoint of $\mathbf{m}$?\ 

\medskip
If ${\mathbf t}$ and ${\mathbf m}$ are disjoint then $\xi$ will linearly increase or decrease,
depending on the relative position of ${\mathbf t}$ and ${\mathbf m}$ on a~real axis.\
We will finally get either $d\,({\mathbf t}, \overline{m}) = \xi + \textrm{w}({\mathbf m})$ or
$d\,({\mathbf t}, \overline{m}) = \xi - \textrm{w}({\mathbf m})$ (this must be a~positive number
as\ ${\mathbf t}\,\cap\,{\mathbf m}=\varnothing$).\ Consequently we have bounded
the true distance $d\,({\mathbf t}, m^{\star})$ with uncertainty
$\left| d\,({\mathbf t}, \overline{m}) - d\,({\mathbf t}, \underline{m})\right|=\textrm{w}({\mathbf m})$.\
This is indeed very remarkable result, especially when compared with the so called
natural interval extension of a~basic building block of $\chi^{2}$, namely the expression
`${\mathbf t}-{\mathbf m}$'.\  There we always have $\textrm{w}({\mathbf t}-{\mathbf m}) =
\textrm{w}({\mathbf t}) + \textrm{w} ({\mathbf m}) > \textrm{w}({\mathbf m})$.\
 In our approach the uncertainty of
a~distance in question never exceeds the uncertainty of an individual measurement
and -- surprisingly -- it does {\em not\/} depend on current accuracy of unknown parameters.\ 

\medskip
It remains to show how this result changes when ${\mathbf t}$ and ${\mathbf m}$
overlap.\  Previously, thanks to the disjointness of ${\mathbf t}$ and ${\mathbf m}$,\
only one argument of (\ref{H-point}) was `active' at all times, meaning that its value
defined the distance.\ Now, at some point $m^{\prime}$ (not necessarily $m^{\star}$), the two
arguments can exchange their roles and the other one may become `active'.\  If this
happens (it may not) then the direction of change of\  $\xi$\  will change too,
thus `un-doing' the already acquired change.\  Therefore the final change of\
$\xi$\  cannot even reach $\textrm{w}({\mathbf m})$.

\CCLsection{Main result}
The analysis just presented is intuitive and easy to follow, but still crude.\  
It says nothing about the values of bounds, being limited only to their separation.\
This is not enough to be useful in practice.\
In particular, it is easy to show that $m^{\prime}$ at which the two arguments
of (\ref{H-point}) are equal to each other is $m^{\prime}=\left(\underline{t} + \overline{t}\right)/2$
-- the center of interval ${\mathbf t}$\  (but only when\ $m^{\prime}$\ also belongs to\ 
${\mathbf m}$,\ otherwise there is no such point and switching of roles does not occur).\
For such $m^{\prime}$, the distance $d\,({\mathbf t}, m^{\prime}) = (\overline{t}-\underline{t})/2$.\
With this fact in mind, we are able to derive the exact bounds for the distance $\rho$,\
between a~predicted value ${\mathbf t}$ of an~experimental outcome, under known
environment ${\mathbf x}$, and the true value $m^{\star}\,\in\,{\mathbf m}$, in the same
circumstances ${\mathbf x}$.\  They are following:

\begin{itemize}
\item when $\textrm{c}({\mathbf t})\,\in\,{\mathbf m}$:
\begin{equation}
\begin{tabular}{lccl}
lower bound: &  $\underline{\rho}$ &=& $\frac{1}{2}\textrm{w}({\mathbf t})$\\
&&&\\
upper bound: &  $\overline{\rho}$ &=& $\max\,\left[\,d\,({\mathbf t}, \underline{m}),\,
 d\,({\mathbf t}, \overline{m})\,\right]$
\end{tabular}
\label{fin-a}
\end{equation}

\item  when $\textrm{c}({\mathbf t})\,\not\in\,{\mathbf m}$:
\begin{equation}
\begin{tabular}{lccl}
lower bound: &  $\underline{\rho}$ &=&  $\min\,\left[\,d\,({\mathbf t}, \underline{m}),\,
 d\,({\mathbf t}, \overline{m})\,\right]$\\
&&&\\
upper bound: & $\overline{\rho}$ &=& $\max\,\left[\,d\,({\mathbf t}, \underline{m}),\,
 d\,({\mathbf t}, \overline{m})\,\right]$,
\end{tabular}
\label{fin-b}
\end{equation}
\end{itemize}
where $\textrm{c}(\cdot)$ stands for the center of its interval argument, here\
$\textrm{c}({\mathbf t})=\frac{1}{2}\left(\underline{t}+\overline{t}\right)$, and
$d\,(\cdot, \cdot)$ is a Moore-Hausdorff distance between intervals.\
Please note that generally
\begin{equation}
d\,({\mathbf t}, {\mathbf m})\,\ne\,\max\,\left[\,
d\,({\mathbf t}, \underline{m}),\, d\,({\mathbf t}, \overline{m})\,\right],
\label{unlabelled}
\end{equation}
so the above bounds cannot be written in a~more compact form.

In addition, we have introduced a new symbol, $\rho$, for the mathematically correct
distance between the true measured quantity and its interval theoretical estimate.\
${\boldsymbol\rho}$ stands for its interval enclosure.\ 
The intuitively shown relation\ $\textrm{w}({\boldsymbol\rho})\,\leqslant\,\textrm{w}({\mathbf m})$\
remains true.

\CCLsection{Discussion}
In conclusion, the recommended form of interval version of $\chi^{2}$ functional is:
\begin{equation}
{\boldsymbol\chi}^{2} = \sum_{j=1}^{N} \left[ \frac{ {\boldsymbol\rho}\, ({\mathbf t}_{j}\,
({\mathbf p}_{1}, \ldots, {\mathbf p}_{k}),\
{\mathbf m}_{j})} {\textrm{w}({\mathbf m}_{j})} \right]^{2},
\label{chi-int}
\end{equation}
where ${\boldsymbol\rho}=\left[\,\underline{\rho}, \overline{\rho}\right]$ is given by
(\ref{fin-a}) and (\ref{fin-b}), respectively.\  To recover the unknown parameters
${\mathbf p}_{1}, \ldots, {\mathbf p}_{k}$, one has to find a~global minimum of (\ref{chi-int})
with respect to those parameters.\  This task may be accomplished using procedures
developed mostly by Luc Jaulin and \'Eric Walter (notably SIVIA algorithm, set-inversion
methods) as well as ideas first put forward by Shary \cite{RC7-497}.\ No matter which
approach will be adopted, it is clear that no crisp values can be ever obtained, just
because all ${\mathbf m}$'s are intervals.\  On the other hand, this very feature is
highly desirable, since the uncertainties of searched parameters are evaluated
very credibly and as nearly a~side effect, with no extra effort.\ This observation
sheds new light not only on the problem of experimental data fitting in general, but
also substantially changes our perspective on reliable estimates of uncertainties
in indirect measurements.

\medskip
Once the measurements are completed, the values of all ${\mathbf m}$'s, as well as their
widths, are fixed.\  This makes possible to use widths of measurements as the natural
unit lengths in every direction of $N$ possible.\  That is why ${\mathbf m}$'s are
present in denominators of all components of the sum (\ref{chi-int}).\ 
When we speak of distances this is the only choice of appropriating individual weights
to all measurements.\  There is no space left for arbitrariness, as it sometimes happens
in other versions of the so called {\em weighted LSQ regression\/}.

As the measurements are fixed during calculations, the interval enclosure
${\boldsymbol\rho}$ of the distances between the true unknown values $m^{\star}_{j}$
and their theoretical predictions ${\mathbf t}_{j}$, is perfect and cannot be further
improved.\ Moreover, those widths are uniquely determined by the accuracy of
the real measurements, not guessed or necessarily subjectively evaluated
by a~human expert.\ The optimal widths are very fortunate, since the lower is the
width of the interval extension of a~function being minimized, the more precisely the
global minimum may be located.\  At least at this respect our procedure definitely
beats the natural approach.

\medskip
In the second part of this work a connection between interval hulls of solutions
and statistical description of their uncertainties will be demonstrated and discussed.

\CCLsection*{Acknowledgments}
This work was done as a~part of author's statutory activities at  the  Institute  of Physics,
Polish Academy of Sciences.

\end{document}